# Observing the crack tip behaviour at the nanoscale during fracture of ceramics

*Oriol Gavalda-Diaz*[1,2,*], *Max Emmanuel*[1], *Katharina Marquardt*[1,3], *Eduardo Saiz*[1], *Finn Giuliani*[1]

[1]Department of Materials, Royal School of Mines, Imperial College London, London, SW7 2AZ, UK

[2]Composites Research Group, Faculty of Engineering, University of Nottingham, Nottingham, NG7 2GX, UK

[3]Department of Materials, University of Oxford, Parks Road, Oxford, OX1 3PH, Oxfordshire, UK

*Ultimately, brittle fracture involves breaking atomic bonds. However, we still lack a clear picture of what happens in the highly deformed region around a moving crack tip. Consequently, we still cannot link nano to atomic-scale phenomena with the macroscopic toughness of materials. The unsolved challenge is to observe the movement of the crack front at the nanoscale while extracting quantitative information. Here we address this challenge by monitoring stable crack growth inside a TEM. Our analysis demonstrates how phase transformation toughening, previously thought to be effective at the microscale and above, promotes crack deflection at the nano-level and increases the fracture resistance by ~50%. The work will help to connect the atomistic and continuous view of fracture in a way that can guide the design of the next generation of strong and tough materials demanded by technologies as diverse as healthcare, energy generation or transport.*

## 1. Introduction

The structural capabilities of brittle materials such as ceramics or refractory metals are often limited by their poor fracture resistance. Not surprisingly, a large effort has been placed on increasing their toughness to meet the needs of a broad range of applications where their thermochemical resistance offers significant advantages[1–3]. Multiple strategies have been implemented. They range from the transformation toughening exhibited by the zirconia ceramics used in orthopaedic implants[4] to crack bridging and fibre pull-out in ceramic matrix composites for nuclear and aerospace applications[5,6]. As the demands of the applications are continuously growing so does the need for even stronger and tougher materials[7]. These developments have been accompanied by systematic studies on the fundamental mechanisms that determine fracture resistance and the ways to quantify it experimentally. However, and despite the maturity reached by the field, there is still a very significant gap between the well-established continuous linear-elastic analysis of toughness and the atomistic models of crack propagation. This gap limits our fundamental understanding and hampers practical progress.

The continuous linear elastic fracture mechanics (LEFM) analysis results in a stress singularity at the crack tip and goes around this issue by defining a "process zone" (with a typical size of few nanometres and below) around this tip where the behaviour is no longer linear-elastic. It defines toughness as a critical stress concentration ($K_c$) that has to be reached for the crack to propagate. $K_c$ determines the stress distribution away of the process zone at this critical point and can be linked to the fracture energy, $G_c$, through a very simple relationship ($G_c=K_c^2/E$, where E is the Young's modulus). Note that as reported by Issa et al.[8] plane stress conditions prevail in our tests as the sample thickness is similar in magnitude to the expected plastic zone ($r_p$) in a stiff and brittle ceramic. However, at the atomic scale bond breaking at the crack tip happens within the process zone. According to the simplest of atomic views, the fracture energy should be equivalent to the one needed to create two new surfaces ($2\gamma_s$) but macroscopic experimental measurements are usually well above. This observation supports the accepted view that common toughening mechanisms in brittle materials act at the microscale by shielding the crack tip from the applied stresses and that some of these mechanisms could be detrimental to the strength of the material[2]. Unfortunately, we still lack the comprehensive picture of how the fracture resistance develops from the atomic to the nanoscale and up. Furthermore, there is evidence suggesting that toughening may also occur at these scales and atomic models predict a series of phenomena such as lattice trapping[9], void formation ahead of the crack tip[10–12] or plasticity in the process zone[13] that are very challenging to verify experimentally. To support the rational design of the next generation of structural materials able to combine high strength and toughness we need to address these open fundamental questions to build a holistic understanding of crack propagation and the generation of fracture resistance.

In recent years we have seen a very significant progress on the development straining experiments from the micron to the nanoscale coupled with in-situ high-resolution imaging[14], This includes the development of new TEM in-situ setups capable of linking the performance of materials to atomic and nanoscale events such as the role of interfaces in plastic deformation[15] or the effect of ion intercalation in energy storage devices[16]. These set ups create new opportunities to observe the materials response close to a crack tip. The results show that while in some materials the fracture energy at the microlevel approaches the expected theoretical value of twice the surface energy (once extrinsic toughening mechanisms acting at larger length scales are eliminated)[17], in others this is not the case[18,19]. In this respect, in-situ nanomechanical testing in the transmission electron microscope (TEM) offers the possibility to observe crack propagation at length scales that start to approach those accessible in atomistic simulations. Further developments in this direction should enable us to link directly model and observation and obtain quantitative data that combined with experiments at larger length scales can feed the much-needed comprehensive multiscale failure models. Several in-situ nanomechanical testing setups in the TEM have been developed with the aim of observing and quantifying the plastic deformation that brittle materials experience at the nanoscale by either using tension[20–24], shear[23] and compression[23,25,26] setups. Fracture indentation[27–29] and single cantilever bending[30] tests have also been used to propagate cracks and evaluate their fracture morphology at this scale. Nevertheless, in these setups, cracks in brittle materials propagate in an unstable manner. Therefore, it is possible to observe the material immediately before and immediately after fracture, but the information that can be extracted of the atomic and nanoscale events occurring during the progression of a crack front is very limited. To address this challenge, we are proposing an approach capable of recording nanoscale dynamic events during crack propagation.

Being inspired by the original work from Obreimoff[31] and the later adaption of the test to SEM[17] and TEM[32] in-situ setups, we are proposing an optimized wedge-driven Double Cantilever Beam (DCB) test which achieves stable crack growth inside a TEM when testing stiff and brittle materials. This setup helps us to enhance our mechanistic understanding of failure by resolving in-situ the nanoscale structure of the crack front during propagation. We use this technique to study the fracture of two model brittle materials, SiC, the base of modern tough ceramic matrix composites, and $ZrO_2$, the poster material for transformation toughening used now in high-end applications such as orthopaedic implants or thermal barrier coatings. In the later material, our in-situ set-up enables something that has been considered "impossible or really difficult"[33]: resolving the transformation toughening at the crack tip. Our work shows how stress shielding happens during early stages of crack propagation and reports the toughening mechanisms active at the single grain level. Our observations show how a stress-activated phase transformation occurs well below the microscale, challenging the rules of brittle crack growth.

2. Results

**2.1 In-situ stable crack growth at the nanoscale**

We use a TEM in-situ setup to visualise the nanoscale events occurring when cracks propagate in brittle materials. The notable feature of this setup is that we can achieve stable crack growth even when testing brittle and stiff materials such as SiC or $ZrO_2$. The stable nature of the crack allows us to use the wide range of analytical techniques available in a TEM to describe the nanoscale structure around a crack tip. This has been achieved by redesigning the wedge-driven DCB test traditionally used to measure fracture properties of materials at the macroscale and more recently implemented in SEM and TEM in-situ tests[17,32]. In our setup we propose a method which uses relatively large samples (approx. 2x2x0.1 mm), and in which nano-DCB samples are prepared using Focused Ion Beam (FIB). The DCBs contain two thick arms (approx. 1x1 µm) to facilitate the position of the wedge and reduce the through thickness compliance. An electron transparent region (approx. 1x0.1 µm) is milled at the centre to observe the fracture process at the nanoscale (**Figure 1**). The opening displacement of the arms is applied with a custom-built diamond wedge with an angle of ≈ 65°. Overall, the reduced compliance of the setup allows us to stabilise the fracture test and to resolve the fracture process at the nanoscale (**Figure S1**). We can also pause the test at a given opening displacement and analyse the crack tip with the analytical techniques available in the TEM. More details on the sample preparation and testing conditions are given in the **Methods** section in the *Supplementary information*. In this work we show an example of how we combine TEM imaging and diffraction to explain the fracture process of SiC and $ZrO_2$ based materials.

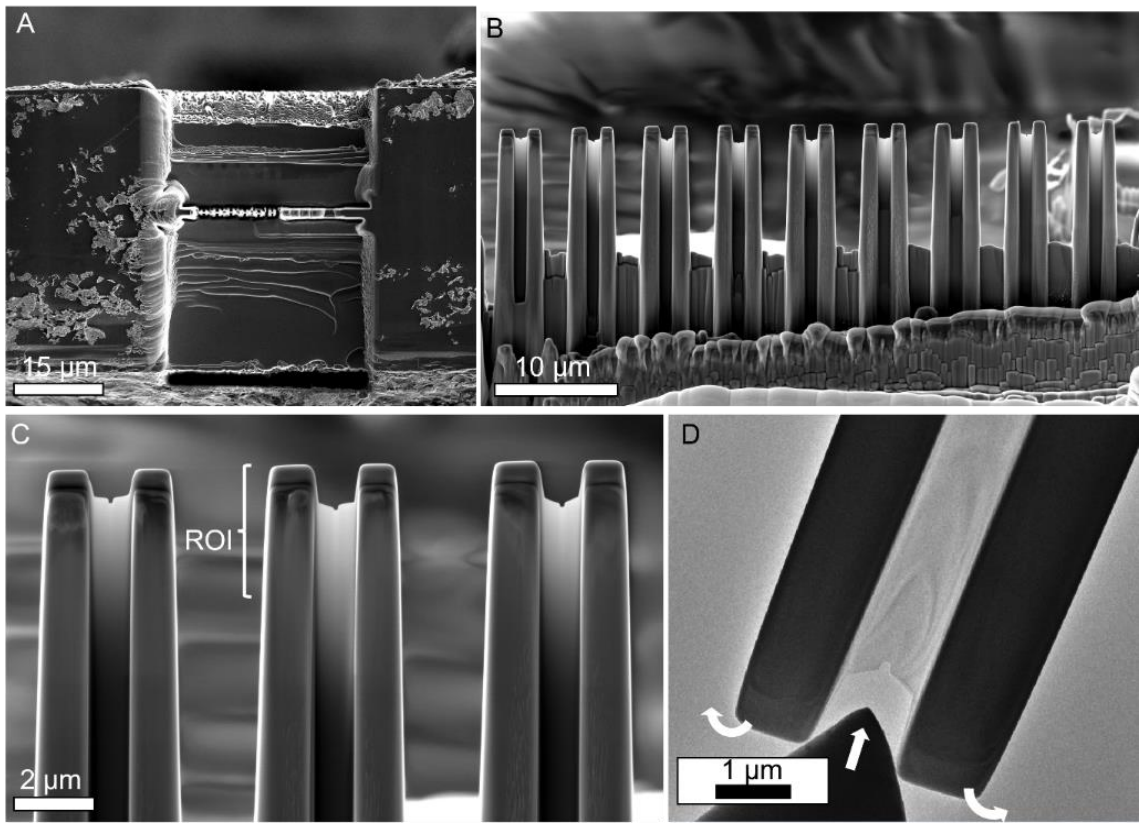

*Figure 1 Nanofabricated single crystal Double Cantilever Beams (DCBs). (A) Top view of the sample after being thinned. (B) Front view of several DCBs. (C) Close up of a DCB*

*showing the two arms (approx. cross section of 1x1 µm) and the thinned region (approx. 1x0.1 µm) with a FIB-milled notch. The region of interest (ROI) shows the length of sample used for crack growth (≈ 1-2 µm). (**D**) Wedge-actuated DCB test showing early stages of the loading and the bending contours around the notch.*

## 2.2 Crack tip behaviour during brittle fracture

Our stable crack growth experiments allow us to observe how a crack tip moves within a "model" brittle ceramic material such as hexagonal SiC. **Figure 2A** shows a frame obtained during the stable crack propagation experiment splitting the basal plane (see also **Figure S2** and *Movie 1*). By using the TEM micrograph frames obtained during the in-situ experiment we get the near crack tip profile by measuring the crack opening (2u) at different distances from the crack tip (x). Based on Irwin's equation we can link the near tip profile to the toughness of the material by using **Eq. 1**[34,35].

$$u(x) = \frac{K_{ic}}{E}\sqrt{\frac{8x}{\pi}}$$  **Eq. 1**

In this case, we are simplifying the analysis by not taking into consideration the anisotropy of the crystal and considering a Young's modulus of $E_{SiC}$ = 480 GPa. We use two values of toughness $K_{Ic}$ = 3.3 and 2.18 MPa.m$^{1/2}$ which represent the literature data available for SiC at the macroscale and twice the surface energy obtained with DFT simulations[17], we plot **Eq. 1** together with our experimental values (see **Figure 2B**). The error analysis coming from the measurement of the crack opening from the TEM frames has been taken into consideration in the experimental data points and has been detailed in the *Supplementary Information* (**Figure S3** and **S4**). Our nanoscale fracture data is in good agreement with the fracture toughness found at the micron scale and suggests that the fracture energy values (i.e., toughness) are close to the twice surface energy (2γ$_s$) of the material[17] as expected from an ideal brittle fracture.

As early suggested by Irwin, LEFM builds on the assumption that in brittle materials a small process zone forms ahead of the crack tip. This is needed to address the stress singularity generated at this tip by the continuous analysis. The different theories agree that when the process zone is generated by plasticity, its extension depends on the toughness ($K_{Ic}$) and yield stress (σ$_y$) of the material. In general, it is well established that the size is proportional to ($K_{Ic}$/σ$_y$)$^2$ and in ceramics is expected to be of the order of nanometres.

Our test allows us to monitor the nanoscale events happening at the crack tip when fracturing SiC. Differently to a post-mortem analysis, this is done while the wedge is inserted with the arms bended. Therefore, we can observe the zone around the crack tip in-situ when a load is applied. **Figure 2C** shows the crack tip behaviour when growing a crack for ≈ 40 nm along the basal plane. There is a damaged area ahead of the tip and which resembles the process zone shape expected for a Mode I fracture test[36]. Based on our fracture toughness data we use a value of $K_I$ = 2.18 MPa.m$^{1/2}$ to plot in **Figure 2D** the theoretical process zone overlapped with the damaged area in our experiments (see **Eq. S1** in the *supplementary information*) for three arbitrary values of yield stress that give a good correlation with our experimental data. In SiC, a process zone with a radius between 30-50 nm suggests a yield stress ranging between 4-6 GPa. Literature in 6H SiC reports a critical resolved shear stress (CRSS) for basal and prism slips of 5-6 GPa[37]. Approximately, we can assume that $\sigma_y \approx 0.5\ CRSS$. Our larger values ($\sigma_y$ = 4-6 GPa) suggest that plasticity might also be active in other crystallographic planes. These observations support the view that in brittle ceramics such as SiC the process zone is generated via plastic deformation.

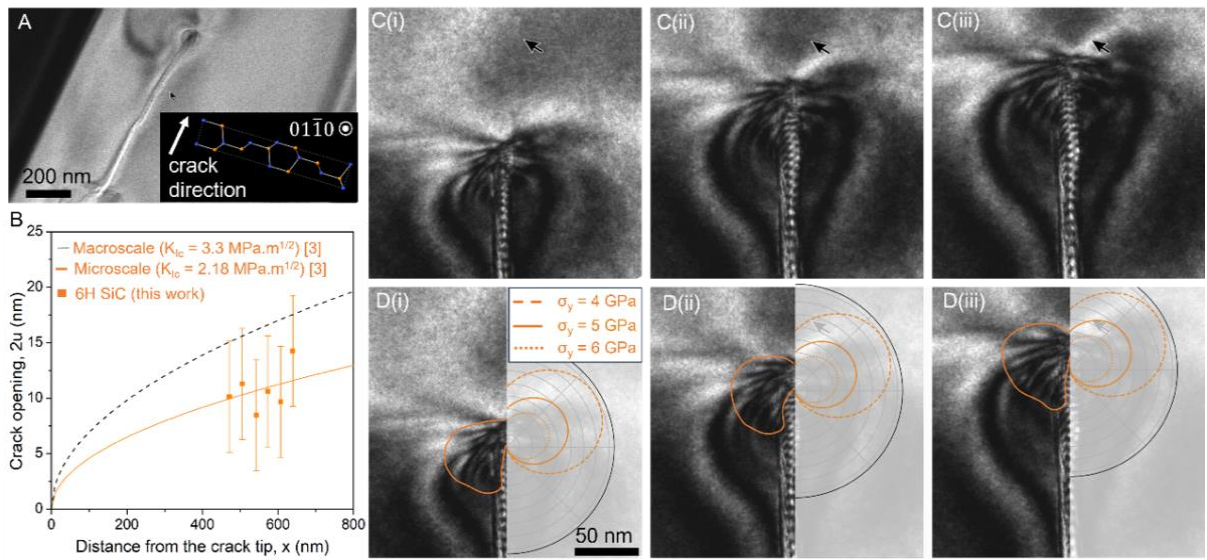

*Figure 2 Stable crack growth in SiC. (**A**) Lower magnification images of the crack. (**B**) Crack openings (2u) at different distances from the crack tip overlapped with the expected values of toughness at the micron and macroscales. (**C**) High magnification frames showing the progress of the crack tip (Movie 1). (**D**) Frames overlapped with the predicted plastic zone for different values of yield stress (σ$_y$). The contour used to highlight the experimental process zone is just an approximation to guide the eye of the reader.*

## 2.3 Toughening of ceramics at the nanoscale

To observe toughening at the nanoscale, we select ZrO$_2$ due its exceptional capability for the customisation of fracture resistance. This is achieved by stabilising the tetragonal phase at room temperature using low contents of an oxide, commonly yttria or ceria. The tetragonal phase transforms to monoclinic in the presence of stress, for example at the crack tip. This transformation is diffusionless and of martensitic nature and induces compressive residual stresses ahead of the crack and can result in a 3 or 4 fold increase in fracture toughness at the macroscale[38,39]. In the case of Yttria Stabilised Zirconia (YSZ), low contents of yttria (3-8 mol%) stabilise the tetragonal phase with a high toughness. When stabilising the cubic phase (8-12 mol%) the toughness increase is lost. In this work we use single crystals of yttria stabilised zirconia with 4 and 9 mol% (4YSZ and 9YSZ) to follow the nanoscale response of the tetragonal and cubic polymorphs respectively.

DCB tests were performed splitting ≈ (001) planes in both samples. In the case of 9YSZ, straight cracks (with an absence of deflection and twisting) are systematically observed and highlight the brittle nature of the material (see **Figure 3A** and *Movie 2*). High-resolution TEM images and diffraction studies confirm the lack of phase transformation and just a process zone of ~ 10-20 nm ahead of the crack tip (**Figure 3B**). As in the case of SiC, the size of the process zone is consistent with theoretical predictions of plasticity.

When testing 4YSZ we observe how cracks follow a more tortuous path with evidence of crack deflection and twisting (see **Figure 3C** and *Movie 3*). Similar crack growth mechanisms were observed in all the tests (**Figure S5**). This suggests that the crack is being redirected by the local nanostructure formed ahead of the tip. It is remarkable to observe how deflection and twisting mechanisms, common in bulk fracture testing of polycrystals yttria stabilised zirconia[39], arise at the nanometre scale within a single grain.

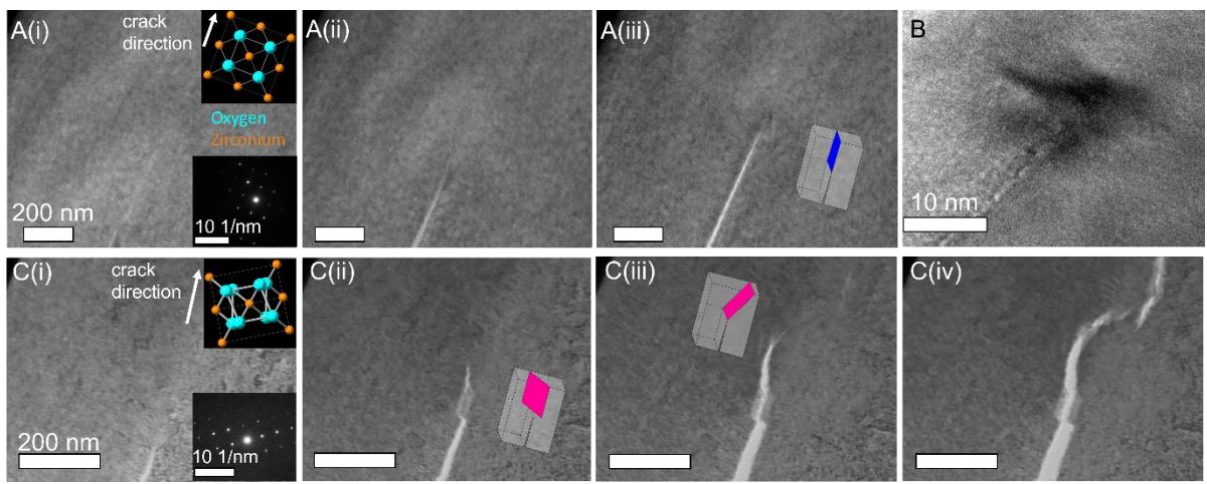

*Figure 3 Stable crack growth in $ZrO_2$. (A) Splitting the ≈ (001) plane in the cubic polymorph (9YSZ) and showing how the crack propagates straight without deflection/twisting (Movie 2). (B) High magnification image of the crack tip in 9YSZ, showing the lack of phase transformation and just the plastic zone at the crack tip. (C) Splitting the ≈ (001) plane in the tetragonal polymorph (4YSZ) highlighting the crack tilting and twisting (Movie 3).*

By following the same approach as with SiC, we use the crack opening to measure the toughness of ZrO$_2$ at the nanoscale (**Figure 4A**). Considering a Young modulus of E$_{4YSZ}$ = 214 GPa and E$_{9YSZ}$ = 220 GPa[40] (similar to the ones reported in micromechanical testing[41]), our data suggests approximate values of nanoscale fracture toughness of 1.8 and 2.7 MPa.m$^{1/2}$ for 9YSZ and 4YSZ respectively for cracks lengths shorter than 1 µm. Note that for the 4YSZ samples we analyse cracks with minimal deflection to ensure an accurate measurement of the crack tip distance (x). This also highlights that the increase in toughness reported here is mainly dictated by the phase transformation and not due to an increase of crack surface.

**Figure 4B** compares the values of fracture toughness that we measure for single crystals at the nanoscale with average literature data for polycrystalline YSZ tested at the macroscale[38]. 9YSZ shows good agreement with the values measured in macroscopic cubic polycrystalline YSZ samples, highlighting its brittle nature and absence of active extrinsic toughening mechanisms in the material. In the case of 4YSZ, we measure lower values compared to the reported ones for large Tetragonal Zirconia Polycrystalline (TZP) samples but around 50% larger than for 9YSZ. This highlights how the transformation zone and crack shielding can be dependent on the following factors:

(i) Length scale: When tested at the macroscale in vacuum, TZP displays a rising R-curve[42], suggesting that the transformation zone ahead of the crack tip increases with crack length. Typically, the post-mortem analyses of macroscopic samples show fully transformed grains around the crack. Our data shows how toughening starts at very short crack lengths and how the toughening mechanisms are activated within a single grain with a transformation zone that forms and increases in size after the crack propagates for few nanometres.

(ii) Sample geometry: In our work we propagate cracks within an electron transparent region, removing part of the 3D nature of the transformation zone and also limiting the coincidence planes in which tetragonal to monoclinic phases can nucleate.

(iii) Microstructure: Within a polycrystalline material, random grain orientations might produce transformation zones with more complex residual stress fields and more tortuous crack paths.

Overall, with our work we highlight that the toughening mechanisms traditionally reported in zirconia can already be active in early stages of crack propagation and within a single grain. By doing the experiment inside a TEM we can also explore the nanoscale events dominating this toughening mechanism.

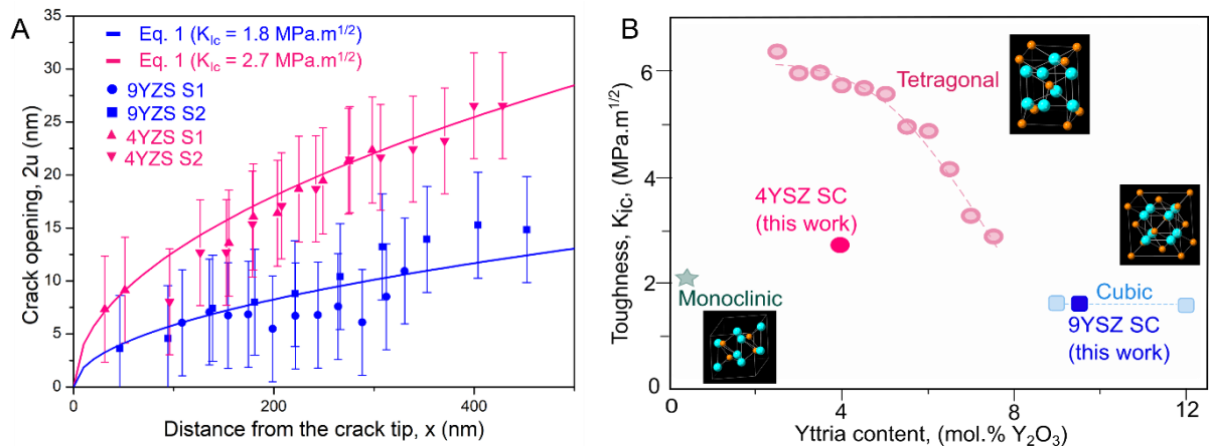

*Figure 4 Fracture toughness of zirconia at the nanoscale. (**A**) Crack openings (2u) at different distances from the crack tip (x) for 4YSZ and 9YSZ and the calculated $K_{IC}$ obtained by fitting Eq. 1. (**B**) Average literature toughness values for the different polymorphs of polycrystalline YSZ at the macroscale (data extracted from[38]) overlapped with our data of toughness at the nanoscale.*

## 2.4 Transformation toughening at the crack tip

The transformation toughening leading to the tortuous crack path in tetragonal zirconia is studied more in depth by performing tests at higher magnifications and highlighting diffraction contrast with a smaller objective aperture in the TEM. When the crack tip advances, we observe how darker regions form ahead of the crack tip (**Figure 5A**). These darker regions nucleate in the crack front and suggest that a stress-activated tetragonal to monoclinic phase transformation is progressing (see *Movie 4*).

**Figure 5B** shows the crack in **Figure 5A** after being fully unloaded and reloaded (*Movie 5*). See the load-displacement curve for the first and second loading in **Figure 5C.** When the load is removed, the dark areas remain present in the sample indicating a stable and permanent nature (**Figure 5B**(i)). During the regrowth of the crack (**Figure 5B**(ii-v)) the extent of the darker areas continue to spread ahead of the crack tip. The shape of the darker area resembles the transformation zone predicted under dilatation and shear stresses in Mode I[39]. Lower magnification images of the extent of the darker regions at different stages of loading and unloading are also shown in **Figure S6**.

The nature of the transformation zone is studied via diffraction and high-resolution imaging. Extra spots in the diffraction pattern (see insert in **Figure 5F**) confirm the presence of the monoclinic phase after crack propagation. The possible monoclinic orientations matching the diffraction patters appears to suffer from a ≈45º rotation from expected lattice correspondence (**Figure S7**)[43]. This is attributed to surface relief caused by the volume expansion in our electron transparent samples.

To qualitatively assess the extent of the transformation process we use Selected Area Diffraction (SAD). We measure diffraction spots corresponding to the monoclinic phase up to ≈ 500 nm away from one the side of the crack, but their intensity decreases when moving further away (**Figure 5I** (i-ii)). At a distance of 300 nm ahead of the crack tip and in the other side of the crack, there is no evidence of phase transformation (**Figure 5I** (iii-iv)). **Figure S8** shows a higher magnification analysis of the lateral extent of the phase transformation. The data suggests that the transformation nucleating at the crack tip (≈ 100 nm) grows in the crack wake during the bending of the beams. Our data also shows the subcritical nature of phase transformation in which remaining tetragonal phase is observed in the transformed regions and imply the coexistence of the monoclinic and tetragonal phases.

Higher magnification TEM images at the vicinity of the crack (**Figure 5G and 5H**) show Moiré fringes which suggest that the monoclinic phase nucleates as twins forming a herringbone structure with monolithic laths within a tetragonal matrix[44]. This is further confirmed by the long streaks observed in the diffraction pattern (insert in **Figure 5F**) and which confirm the 2D nature of the defects.

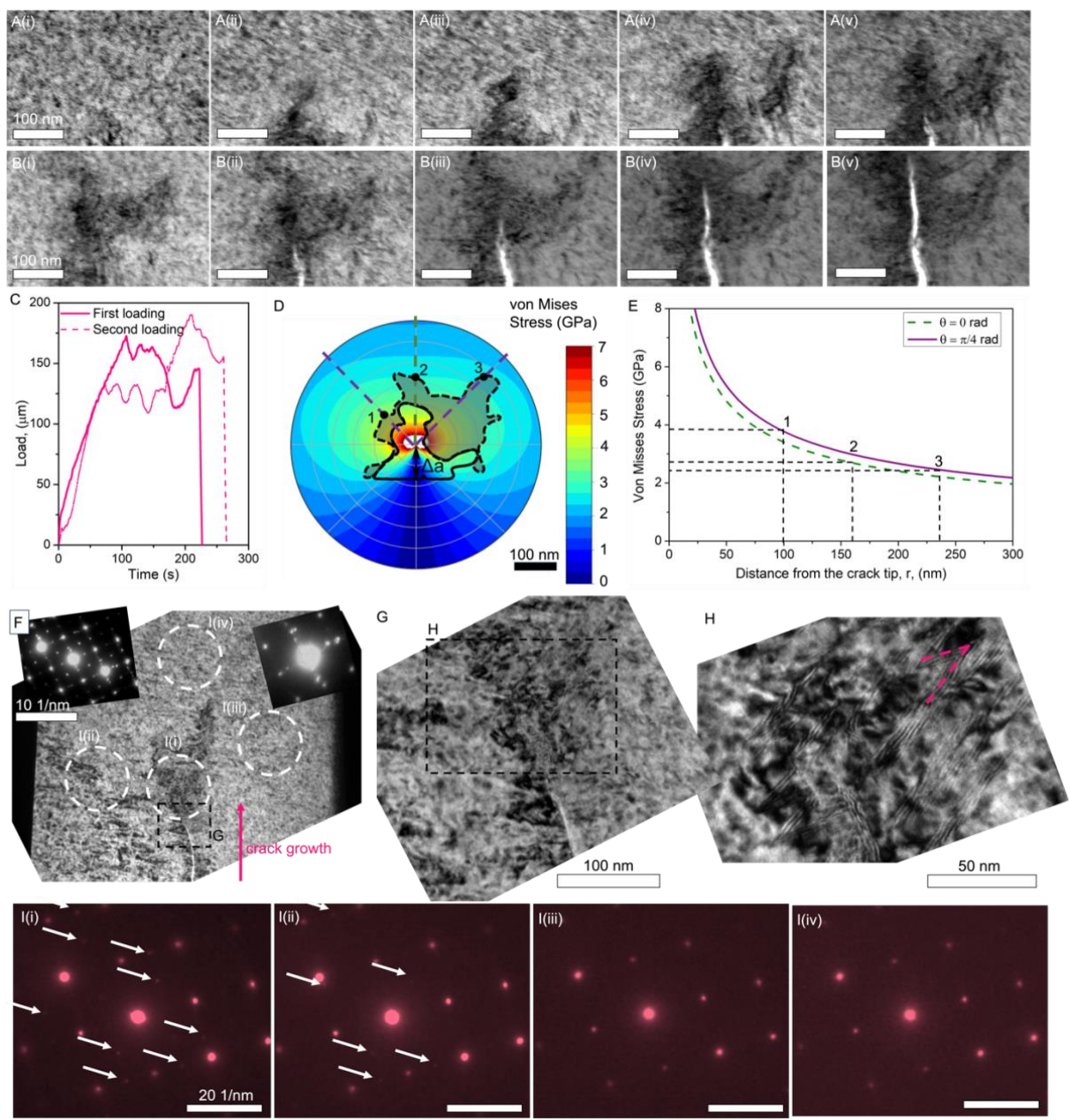

*Figure 5 Transformation toughening in tetragonal zirconia.* (**A**) Darker regions forming ahead the crack tip (Movie 4). (**B**) Regrowth of the crack in A after the sample was unloaded and showing how the darker regions keep nucleating and growing ahead of the crack tip (Movie 5). (**C**) Load-Time curves for the loading and reloading test in A and B. (**D**) Growth of the transformation zone overlapped with the Von Mises stresses with profiles for θ = 0 and 45º in (**E**). (**F**) Low magnification image of the overview of the crack and the four regions used for the SAD measurements. The insets are the DP of the sample after testing. (**G** and **H**) Higher magnification images of the crack showing the formation of moiré fringes and suggesting the formation of a herringbone structure. (**I**) SAD patterns with arrows highlighting the diffraction spots corresponding to the monoclinic phase.

We use the TEM frames to quantify the extend of the transformation zone around the crack tip. The shaded area in **Figure 5D** shows an example of the growth of transformation zone for an increase of crack length (Δa) of ≈ 75 nm (see **Figure S9**). The growth in transformation zone is overlapped with the von Mises stresses predicted via LEFM (see **Figure S10**) when propagating a crack in Mode I and using as stress intensity factor $K_I$ = 2.7 MPa.m$^{1/2}$. To help visualise the stresses reached at the edge of the transformation zone, we plot 2D stress profiles for two given θ angles: 0 and 45º (see green and purple lines in **Figure 5E**). Our results suggest that the critical stress required to trigger the tetragonal to monoclinic phase transformation oscillates between 2 and 4 GPa. These results agree with the yield stress reported for 3YSZ when testing at the microscale using pillar compression[41].

### 3. Conclusions

In this study we have observed in-situ the nanoscale structure ahead of an advancing crack front in different ceramics. By combining different imaging and diffraction techniques, we can explain how this nanostructure is affected by the extreme stress field generated by the crack and how this influences the fracture properties of the material at the nanoscale. We have first shown the progress of a process zone ahead of the crack tip when testing "model" brittle materials such as hexagonal SiC or cubic $ZrO_2$ (9YSZ). Our data suggests that this process zone is generated by plastic deformation and its size agrees with the LEFM predictions. The measured toughness approaches the ideal value of twice the surface energy that could be expected in the small-scale experiments when there is no contribution from large-scale extrinsic toughening mechanisms. From this baseline, we expand our analysis to study the failure process in the presence of transformation toughening in tetragonal $ZrO_2$ (4YSZ). We have shown how we can resolve the stress-induced phase transformation operating ahead of a crack tip and reveal its nanostructure during crack growth. This has given us unprecedented information on how phase transformation changes the local nanostructure and the critical stress required to activate it. Phase transformation at the nanoscale promotes crack deflection within a single grain and raises the fracture resistance for cracks that are only hundreds of nanometres in length.

Overall, this work has shown how visualising the advance of a crack in-situ at the nanoscale could help to unveil the processes active at the tip during propagation while measuring fracture properties at the nanoscale. The goal is to bring the experimental length scales closer to these that can be directly implemented in atomistic models in way that will help us to understand how fracture resistance is generated from the atomic scale and up in brittle materials. This work uses only a small fraction of the possible techniques available in a TEM and other techniques, including spectroscopy or 4D STEM[8], could be explored in order to learn more about the coupling of structure and chemistry during fracture. The final goal is to enhance our fundamental understanding of fracture, linking continuous and atomistic models in a way that will support the design of the next generation of stiff, strong and tough materials.

## 4. Methods

### 4.1 Materials

Three ceramics were used for this study: (i) single crystals of $ZrO_2$ 9.5% $Y_2O_3$ (9YSZ) oriented along the (100) direction supplied by Princeton Scientific Corporation, (ii) single crystals of $ZrO_2$ 4% $Y_2O_3$ (9YSZ) oriented along the (001) direction supplied by Pi-Kem and (iii) single crystals of 6H-SiC oriented along the (0001) supplied by MTI Corporation.

### 4.2 Sample preparation

Both materials were polished down to 50-200 µm thickness using diamond suspensions (6, 3 and 1 µm successively). Afterwards, the samples were placed with the thin edge facing up and the thickness was further reduced locally using a SEM/FIB station (Helios Dual Beam, FEI, US), as shown in **Figure 1A**. The currents used were reduced from 9 nA to 1nA and finishing at 0.3 nA, all at 30kV.

Once the thickness of the sample was reduced to around 3 µm, DCBs were milled with the FIB using 30 kV and 0.3 nA. The centre of the DCB (electron transparent region of around 1 µm in width) was finished at 30 kV and 20 pA. A notch was milled using the same conditions used in the last step. A final cleaning was performed at 5 kV and 20 pA, until the top of the thin area was around 50-100 nm (see **Figure 1C**). EELS was used to confirm the thickness range of the samples.

### 4.3 Tip preparation

A Berkovich indenter commercially available from Hysitron to be used in the TEM in-situ holder was modified to a wedge with the FIB. First, the bottom of the tip was flattened by removing the edge of the Berkovich at 6 nA. Afterwards, the wedge shape was created by placing the tip perpendicular to the ion beam and tilting the tip ± 30°. The final tip geometry had an angle of ≈ 65º and can be seen in **Figure 1D** and **S1B**.

### 4.4 Nanomechanical testing

Fracture tests were performed with a HysitronTM PI 95 TEM PicoIndenter holder, a TEM in-situ mechanical holder. The tests were performed in a FEG TEM (2100F, JEOL, Japan) operating at 200 kV. The displacement rate of the wedge was 1 nm/s, the minimum allowed by the holder to ensure a stable crack propagation. The test was stop when the crack propagated outside the field of view. Crack were propagated within the first 1 -2 µm. The tests were performed at different magnifications to provide an understanding of how crack propagate from the micron to the nanometre scales.

**Acknowledgements**

The authors would like to the EPSRC Future Manufacturing Hub in Manufacture using Advanced Powder Processes, EP/P006566/1.

**Author contributions**

O.G.D., K.M., E.S., and F.G. planned the experiments and designed the data analysis approach. O.G.D. and M.E. performed FIB sample fabrication and performed the experiments. All authors contributed to the final manuscript.